# Revealing the Atomic Structure of Silicate Glasses by Force-Enhanced Atomic Refinement

*Qi Zhou ([1]), Tao Du ([1,2,3]), Lijie Guo ([4]), Morten M. Smedskjaer ([5]), Mathieu Bauchy ([1,\*])*

[1]Physics of AmoRphous and Inorganic Solids Laboratory (PARISlab), Department of Civil and Environmental Engineering, University of California, Los Angeles California 90095, USA
[2]Key Lab of Structures Dynamic Behavior and Control (Harbin Institute of Technology), Ministry of Education, 150090 Harbin, China
[3] School of Civil Engineering, Harbin Institute of Technology, 150090 Harbin, China
[4] National Centre for International Research on Green Metal Mining, BGRIMM Technology Group, 100160 Beijing, China
[5] Department of Chemistry and Bioscience, Aalborg University, 9220 Aalborg, Denmark
\* contact: bauchy@ucla.edu

**Abstract**

Although experiments can offer some fingerprints of the atomic structure of glasses (coordination numbers, pair distribution function, etc.), atomistic simulations are often required to directly access the structure itself (i.e., the positions of the atoms). On the one hand, molecular dynamics (MD) simulations can be used to generate by quenching a liquid—but MD simulations remain plagued by extremely high cooling rates. On the other hand, reverse Monte Carlo (RMC) modeling bypasses the melt-quenching route—but RMC often yields non-unique glass structures. Here, we adopt the force-enhanced atomic refinement (FEAR) method to overcome these limitations and decipher the atomic structure of a sodium silicate glass. We show that FEAR offers an unprecedented description of the atomic structure of sodium silicate. The FEAR-generated glass structure simultaneously exhibits (i) enhanced agreement with experimental neutron diffraction data and (ii) higher energetic stability as compared to those generated by MD or RMC. This result allows us to reveal new insights into the atomic structure of sodium silicate glasses. Specifically, we show that sodium silicate glasses exhibit a more ordered medium-range order structure than previously suggested by MD simulations. These results pave the way toward an increased ability to accurately describe the atomic structure of glasses.

## 1. Introduction and background

Despite their critical role in various technological applications (e.g., optical fibers, display application, nuclear waste immobilization, etc.), the atomic structure of silicate glasses remains only partially understood [1–5]. This is largely due to the fact that, although conventional experiments (e.g., nuclear magnetic resonance, neutron or X-ray diffraction, etc.) offer useful

"fingerprints" of the glass structure (e.g., pair distribution function, coordination numbers, etc.), they do not provide direct access to the atomic structure itself (i.e., the cartesian positions of the atoms). Indeed, although experimental data offer important constraints on the glass structure, these constraints cannot uniquely define the structure itself. For instance, a virtually infinite number of very different structures can present the same pair distribution function [6], so that the full and detailed atomic structure of a glass cannot be uniquely inverted from the mere knowledge of experimental data [7].

As an alternative, complementary route, atomistic simulations offer a direct and full access to the atomic structure of glasses [8–10]. However, atomistic simulations remain plagued by several intrinsic limitations. As a first option, MD simulations are usually based on the melt-quench (MQ) method, wherein a melt is equilibrated at high temperature to lose the memory of its initial configuration and subsequently cooled down to the glassy state [11]. The time-dependent positions of the atoms are then computed by numerically solving Newton's law of motion based on the knowledge of the interatomic forcefield. Although the MQ method mimics the usual experimental synthesis protocol of glasses, most of the MQ-MD simulations are limited to very small timescales (a few ns) due to their computational cost [12]. Hence, MD simulations based on the MQ method are intrinsically limited to cooling rates (typically around 1 K/ps) that are several orders of magnitude higher than experimental ones [9,13–16]. This is a serious drawback for out-of-equilibrium glasses, as their structure and properties depend on the thermal history [17–20]. As such, due to the use of high cooling rates, MQ-MD simulations usually tend to overestimate the glass fictive temperature and, hence, to yield glass structures that are more disordered than those observed experimentally [9,21]. As a second option, RMC simulations have been used to "invert" available experimental data into atomic structures, that is, to generate some atomic structures satisfying given experimental constraints [6,22–24]. However, despite their apparent success (since the structure of RMC-based glasses will necessarily exhibit an excellent agreement with experimental data), RMC simulations are intrinsically based on an ill-defined problem, since, as mentioned above, several glass structure can present the same experimental fingerprint—so that a glass structure cannot be uniquely inverted from experimental data [25]. Altogether, these limitations have raised doubts regarding the ability of atomistic simulations to offer a realistic description of the atomic structure of glasses and, thus, of their properties.

Recently, Drabold *et al.* introduced a new "information-driven" paradigm, wherein the knowledge of both experimental constraints and the interatomic forcefield is simultaneously leveraged to produce realistic atomic models of glasses [26]. Specially, Drabold *et al.* proposed a force-enhanced atomic refinement (FEAR) method, which relies on an iterative combination of RMC refinements (i.e., based on available experimental constraints) and energy minimizations (i.e., based on the knowledge of the interatomic forcefield) [7,26,27]. Unlike alternative hybrid RMC approaches—wherein both structure and energy are optimized simultaneously—the interatomic forces are here only computed during the energy minimization steps, which results in an improved overall computational efficiency [26]. In addition, the FEAR method does not rely on any assumption regarding the weight attributed to structure and energy during the optimization. This method was shown to yield disordered Si and $SiO_2$ atomic structures that simultaneously (i) exhibit an excellent agreement with diffraction data and (ii) are more energetically-stable than those obtained by MD simulations relying on the MQ method [26].

Here, we adopt the FEAR method to refine the structure of a sodium silicate glass—an archetypal model system for more complex, multicomponent silicate glasses [28]. We show that the FEAR method yields a glass structure that simultaneously exhibits enhanced agreement with neutron diffraction data and higher energetic stability as compared to configurations generated by MQ-MD or RMC, while remaining computationally efficient. Based on these results, we find that the medium-range order structure of sodium silicate glasses is more ordered than previously suggested by MQ-MD and RMC simulations.

## 2. Simulations details
### 2.1 Molecular dynamics simulations
To compare the outcomes of MQ-MD, RMC, and FEAR, we simulate a sodium silicate glass $(Na_2O)_{30}(SiO_2)_{70}$ (mol.%). The system comprises 3000 atoms, which has been found sufficient to describe the atomic structure of silicate glasses [29]. We adopt the well-established Teter forcefield, which has been shown to offer a good description of the structural, mechanical, and dynamical properties of sodium silicate glasses [10,28,30–37]. Based on previous studies, a short-range repulsive term is added to avoid the "Buckingham catastrophe" at high temperature [10]. Cutoff values of 8 and 12 Å are used for the short-range and Coulombic interactions, respectively [9,28]. The long-range Coulombic interactions are evaluated with the Particle-Particle Particle-Mesh (PPPM) algorithm with an accuracy of $10^{-5}$ [38]. Note that the PPPM approach relies on the particle mesh method, wherein the particles are interpolated onto a three-dimensional grid. All simulations are carried out using the Large-scale Atomic/Molecular Massively Parallel Simulator (LAMMPS) package [39]. An initial sodium silicate configuration used for the melt-quench method is created by randomly placing the atoms in a cubic box while ensuring the absence of any unrealistic overlap. The system is then melted at 4000 K under zero pressure in the isothermal-isobaric (*NPT*) ensemble for 100 ps to ensure the complete loss of the memory of the initial configuration. The melt is then linearly cooled down to 0 K under zero pressure in the *NPT* ensemble with a cooling rate of ranging from 100 down to 0.001 K/ps. Note that, although these values remain several orders of magnitude larger than experimental cooling rates, the cooling rate of 0.001 K/ps corresponds to a total simulation time of 4 μs, which, to the best of our knowledge, is the longest MD simulation conducted for a silicate glass thus far [9,40]. The use of such a slow cooling rate provides us with a reliable reference atomic structure that can be compared with that yielded by FEAR.

### 2.2 Reverse Monte Carlo simulations
The structure of the same silicate glass is then refined by RMC simulation, as implemented via an in-house fix in LAMMPS [39]. To this end, we use as an experimental constraint the pair distribution function (PDF) obtained from neutron diffraction [41]. The neutron PDF is computed from the partial PDFs $g_{ij}(r)$ as:

$$g_X(r) = \frac{1}{\sum_{i,j=1}^{n} c_i c_j b_i b_j} \sum_{i,j=1}^{n} c_i c_j b_i b_j g_{ij}(r), \qquad \text{(Eq. 1)}$$

where $c_i$ are the molar fractions of element $i$ ($i$ = Si, Na, or O) and $b_i$ are the neutron scattering lengths of the species (given by 4.1491, 3.63 and 5.803 fm for Si, Na, and O atoms, respectively) [42]. Note that, to enable a meaningful comparison between simulated and experimental PDFs, the simulated PDF needs to be broadened, which has been traditionally ignored in previous RMC studies. Here, this is achieved by convoluting the computed PDF with a normalized Gaussian distribution with a full width at half-maximum (FWHM) given by FWHM = 5.437/$Q_{max}$ [43], where $Q_{max}$ is the maximum wave vector used in the diffraction test (here, $Q_{max}$ = 22.88 Å$^{-1}$). The level of agreement between the simulated and experimental PDF is then captured by the $R_\chi$ factor proposed by Wright:

$$R_\chi^2 = \sum_i [g^{\exp}(r_i) - g^{\sin}(r_i)]^2 / (\sum_i g^{\exp}(r_i))^2 \qquad \text{(Eq. 2)}$$

The $R_\chi$ factor is here calculated from $r$ = 0-to-8 Å. The RMC method then consists in the following steps. (i) Starting from a "random" initial configuration, the pair distribution function of the simulated structure is computed and the Wright's coefficient $R_\chi^{\text{old}}$ is calculated (see Eq. 2). (ii) An atom is randomly selected and displaced with a random direction and distance. (iii) The pair distribution function of the new configuration and the new cost function $R_\chi^{\text{new}}$ are computed. (iv) Following the Metropolis algorithm, the new configuration is accepted if $R_\chi^{\text{new}} \leq R_\chi^{\text{old}}$, that is, if the level of agreement between simulated and experimental structure is enhanced by the Monte Carlo move. Otherwise, the atomic displacement is accepted with the following probability or refused otherwise:

$$P = \exp\left[-\frac{R_\chi^{\text{new}^2} - R_\chi^{\text{old}^2}}{T_\chi}\right] \qquad \text{(Eq. 3)}$$

where $T_\chi$ is a constant that controls the probability of acceptance (that is, higher values of $T_\chi$ result in higher probability of acceptance of the Monte Carlo move). Note that the $T_\chi$ constant plays the role of an effective (unitless) temperature, since the term $(R_\chi^{\text{new}^2} - R_\chi^{\text{old}^2})/T_\chi$ is equivalent to the quantity $(U^{\text{new}} - U^{\text{old}})/(kT)$ in the conventional Metropolis algorithm (where U is the energy of the system). Here, the Wright's coefficient $R_\chi$ plays the role of energy (i.e., the metric to be minimized) and $T_\chi$ plays the role of a temperature (i.e., which controls the propensity to accept moves that increase $R_\chi$) [22]. Here, we use $T_\chi$ = 0.001, which was found to result in the lowest final $R_\chi$ upon convergence. Atomic displacements and directions are randomly chosen, with a uniform displacement probability distribution between 0 and 0.2 Å. The simulation box size is kept fixed throughout the simulation at $L$ = 34.47 Å so that molar volume is fixed according to the experimental value of 24.7 cm$^3$/mol [44]. A total number of 500000 RMC moves are attempted until convergence.

## 2.3 Force-enhanced atomic refinement (FEAR) simulations

Finally, we compare for the same sodium silicate composition the outcomes of the MQ-MD and RMC simulations with the results offered by FEAR, as implemented via an in-house fix in LAMMPS [39]. To this end, we start from a "randomized" structure generated by RMC while using a large fictive temperature, namely, $T_\chi$ = 3000. Following the original implementation of the FEAR method [26], the system is then iteratively subjected to a combination of RMC refinements and energy minimizations, wherein each FEAR iteration consists in (i) 5000 RMC steps and (ii) an energy minimization conducted with the conjugate gradient method. As a refinement of the original method [26], we here subject the system to a volume relaxation every 10 iterations, wherein the simulation box is deformed in order to simultaneously reach minimum energy and zero pressure [45,46]. In the following, all properties referring to the FEAR refinement are averaged over 10 independent runs and error bars are evaluated based on the standard deviations. In addition, we dynamically adjust the selectivity of the Metropolis acceptance criterion by linearly decreasing the effective temperature $T_\chi$ from $10^3$ down to $10^{-6}$ during the course of the FEAR refinement. These parameters were found to yield a glass structure exhibiting minimum $R_\chi$ and potential energy values.

## 3. Results and Discussion
### 3.1 Evolution of the structure upon force-enhanced refinement

**Figure 1** shows selected snapshots of the atomic structure of the simulated sodium silicate glass upon FEAR refinement. Overall, we observe that the degree of connectivity increases upon refinement, which manifests itself by the formation of Si–O bonds. **Figure 2** shows the evolution of the neutron PDF of the simulated sodium silicate glass after selected number of iterations of FEAR refinement. The first peak around 1.6 Å arises from Si–O correlations, while the second peak captures second-neighbors (Si–Si and O–O) correlations [47]. We observe that the peak positions of the initial PDF are in fairly good agreement with experimental neutron diffraction data, which indicates that the initial structure exhibits reasonable inter-atomic distances. However, the intensities of the peaks are initially not well reproduced. In contrast, the first peak of the PDF of the intermediate structure (after 30 FEAR iterations) shows an excellent agreement with neutron diffraction data—both in terms of position and intensity—whereas the peaks at larger distance remain less refined. This signals that the refinement of the short-range order occurs faster than that of the medium-range order, likely arising from the fact that refining the short-range order only involves some slight displacements of the neighbors of each atom. In contrast, refining of the medium-range order requires some collective atomic displacements, as well as the breakage and formation of interatomic bonds (which requires larger energy barriers to be overcome). Eventually, we find that the PDF of the final configuration exhibits an excellent agreement with neutron data, both for the short- and medium-range order. Notably, the configuration exhibits a Wright $R_\chi$ factor of about 2.49% (see **Fig. 3a**). Note that a threshold $R_\chi$ factor of 10% is typically used to discriminate "good" from "bad" structures [43].

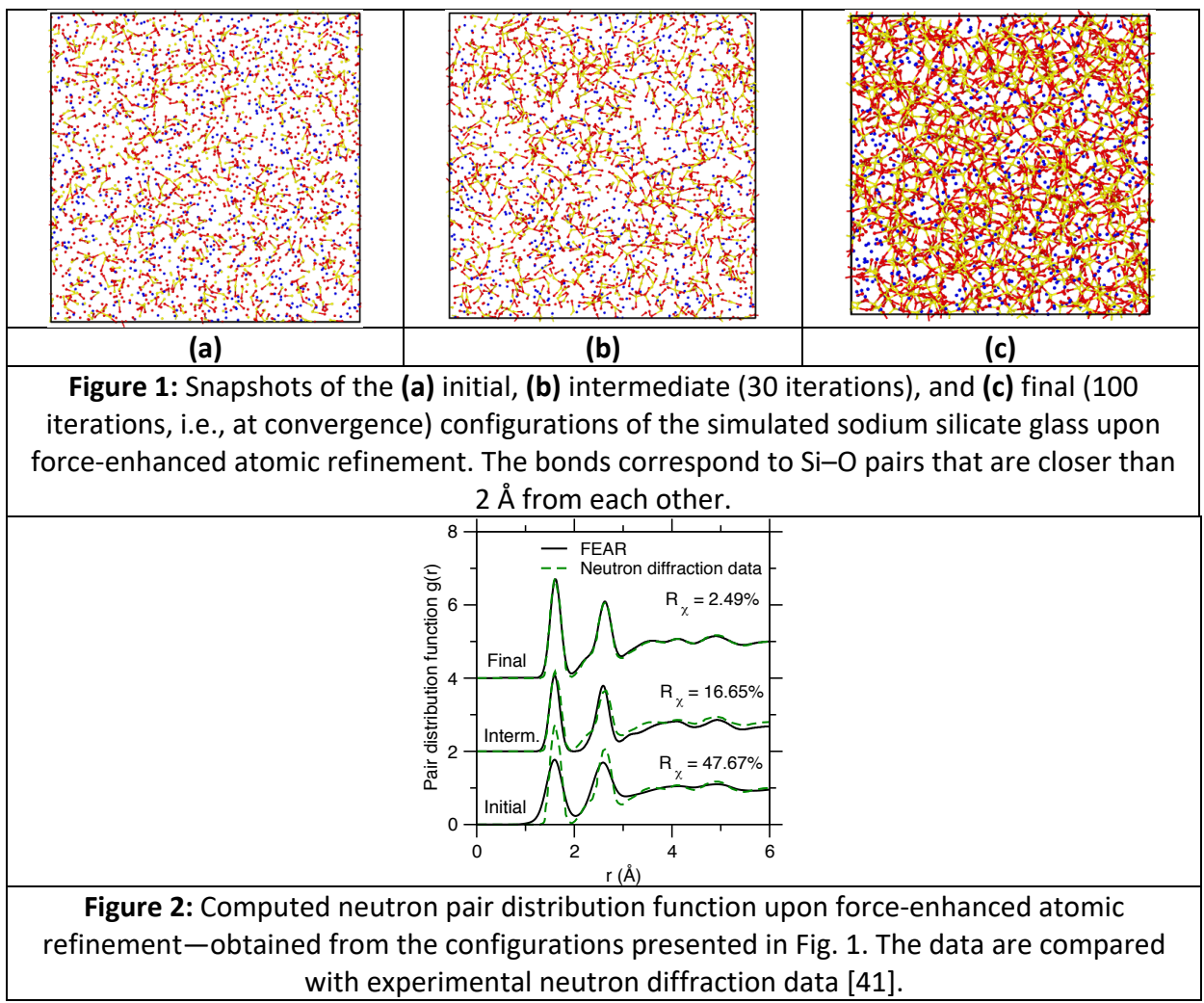

**Figure 1:** Snapshots of the **(a)** initial, **(b)** intermediate (30 iterations), and **(c)** final (100 iterations, i.e., at convergence) configurations of the simulated sodium silicate glass upon force-enhanced atomic refinement. The bonds correspond to Si–O pairs that are closer than 2 Å from each other.

**Figure 2:** Computed neutron pair distribution function upon force-enhanced atomic refinement—obtained from the configurations presented in Fig. 1. The data are compared with experimental neutron diffraction data [41].

*3.2 Comparison with molecular dynamics and reverse Monte Carlo*
We now compare the outcomes of FEAR refinement with those of MQ-MD and RMC simulations. As shown in **Fig. 3a**, we first note that, upon RMC refinement, the $R_\chi$ factor of the simulated sodium silicate glass monotonically decreases and eventually reaches a final $R_\chi$ value of 1.2%, which highlights an excellent agreement with neutron diffraction data (see **Figs. 3b** and **3c**). This is not surprising since RMC refinement solely aims to decrease $R_\chi$. Nevertheless, we note that, although RMC eventually yields a lower $R_\chi$ factor than FEAR, RMC requires a few more iterations than FEAR to converge. This highlights the high computational efficiency of FEAR refinement. In contrast, we note that the MD simulation relying on MQ yields a significantly higher $R_\chi$ value (see **Fig. 3c**). In details, although the glass structure generated by MQ-MD offers an excellent description of the short-range order structure (i.e., first peak of the PDF), we note that MQ-MD does not reproduce well the asymmetry of the second peak. This indicates that the medium-range order of the structure generated by MQ-MD does not fully agree with neutron diffraction data. It is worth noting that the specific value of $R_\chi$ depends on the range of distances used during its calculation, namely, larger distance cutoffs typically results in lower $R_\chi$ values since the PDF necessarily becomes flat and equal to unity at high distance.

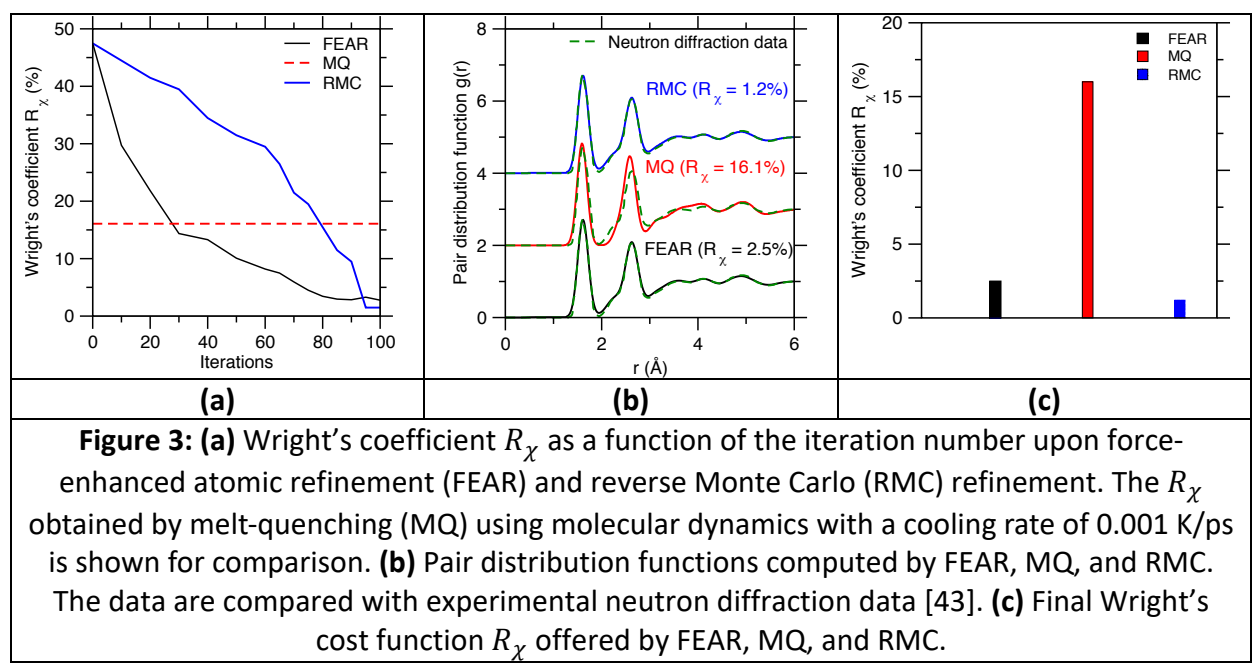

**Figure 3: (a)** Wright's coefficient $R_\chi$ as a function of the iteration number upon force-enhanced atomic refinement (FEAR) and reverse Monte Carlo (RMC) refinement. The $R_\chi$ obtained by melt-quenching (MQ) using molecular dynamics with a cooling rate of 0.001 K/ps is shown for comparison. **(b)** Pair distribution functions computed by FEAR, MQ, and RMC. The data are compared with experimental neutron diffraction data [43]. **(c)** Final Wright's cost function $R_\chi$ offered by FEAR, MQ, and RMC.

To further assess the reliability of the structures predicted by FEAR, RMC and MQ-MC, we now focus on the neutron structure factor, which is computed from the Fourier transform of the neutron pair distribution function (see Ref. [9]). Indeed, although the structure factor contains the same information as the PDF, it places higher emphasis on the structural correlations at longer distances and, hence, can be used to further characterize the medium-range order. In addition, it offers an additional check to assess the quality of the FEAR refinement, since, unlike the PDF, the structure factor is not explicitly taken into account during the refinement.

**Figure 4** shows the neutron structure factor computed from the MQ-MD, RMC, and FEAR configurations. We first find that MD yields a very reasonable description of the structure factor, since all the peaks are fairly well reproduced (see **Fig. 4a**). In detail, we find that the high-$Q$ region of the structure factor ($Q > 4$ Å$^{-1}$) of the glass generated by melt-quenching, which confirms that MD yields a good description of the short-range order of the glass (see **Fig. 3b**). Nevertheless, we observe some discrepancies between neutron diffraction data and MD in the low-$Q$ region of the structure factor. In particular, the sharpness and degree of asymmetry of the first-sharp diffraction peak (FSDP) around 2 Å$^{-1}$ is not well predicted by MD. This signals that MD offers a fairly poor description of the medium-range order structure of the glass.

In contrast, we find that both RMC and FEAR refinements yield a significantly improved description of the low-$Q$ region of the structure factor—albeit to a higher extent in the case of RMC (see **Figs. 4b and 4c**). Importantly, we observe that the position, intensity, sharpness, and asymmetry of the FSDP are very well reproduced by RMC and FEAR refinements. Eventually, we find that RMC and FEAR refinements offer an $R_\chi$ factor (calculated, this time, for the structure factor) of 5.23% and 7.63%, which is significantly improved as compared to that yielded by MD (i.e., 22.16%, see **Fig. 4d**). Again, we find that, although RMC eventually yields a lower $R_\chi$ factor than FEAR, RMC requires a few more iterations than FEAR to converge, which highlights again the high computational efficiency of FEAR refinement (see **Fig. 4d**).

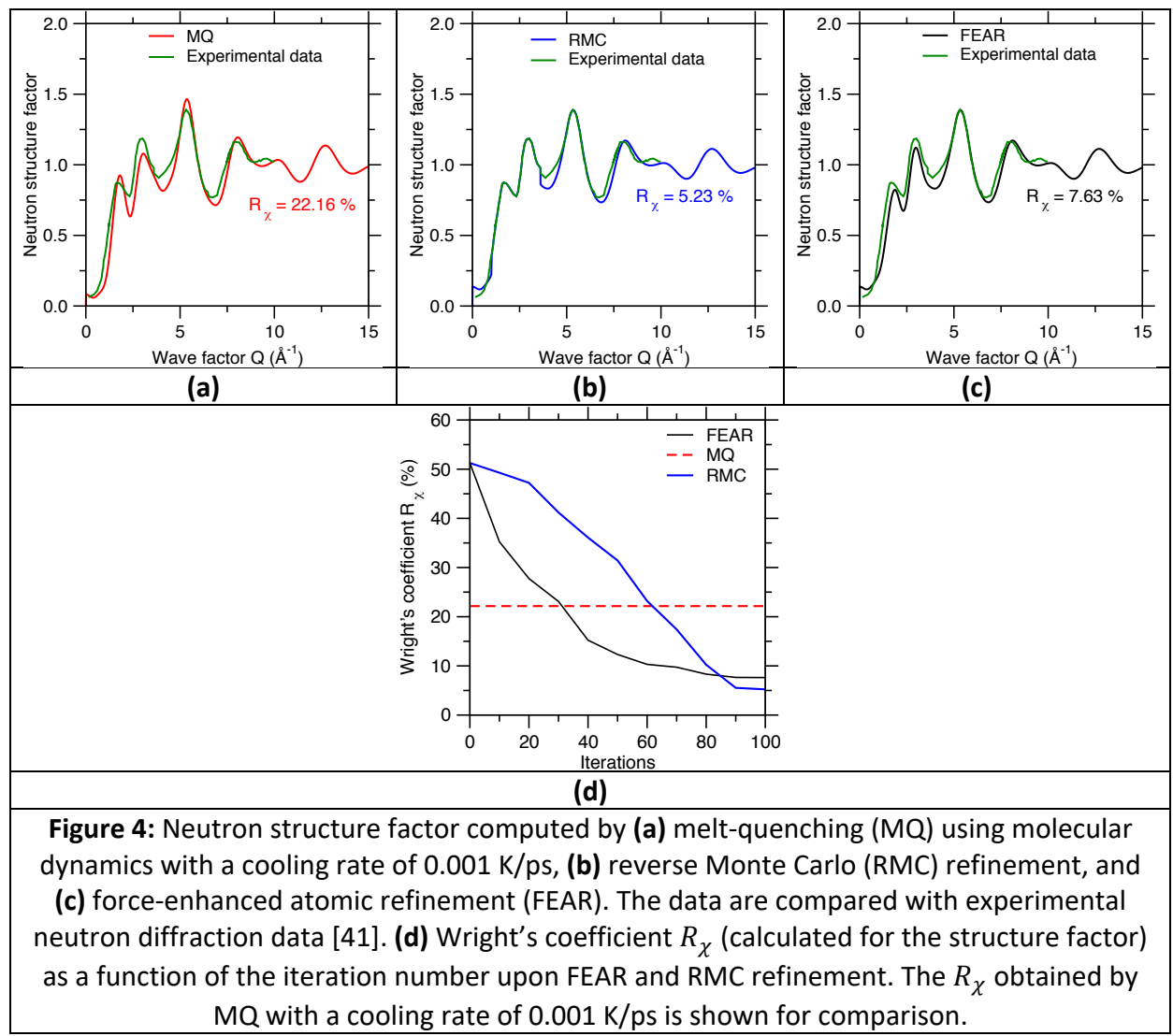

**Figure 4:** Neutron structure factor computed by **(a)** melt-quenching (MQ) using molecular dynamics with a cooling rate of 0.001 K/ps, **(b)** reverse Monte Carlo (RMC) refinement, and **(c)** force-enhanced atomic refinement (FEAR). The data are compared with experimental neutron diffraction data [41]. **(d)** Wright's coefficient $R_\chi$ (calculated for the structure factor) as a function of the iteration number upon FEAR and RMC refinement. The $R_\chi$ obtained by MQ with a cooling rate of 0.001 K/ps is shown for comparison.

We now focus on the thermodynamic stability of the configurations generated by FEAR, RMC, and MQ-MD. **Figure 5** shows the evolution of the molar potential energy of the system upon RMC refinement. Overall, we find that RMC yields a potential energy that is significantly higher (by more than 20 kJ/mol) than that offered by MQ-MD (see **Fig. 5b**). This result is not surprising since RMC is blind to the interatomic energy. Nevertheless, it highlights the fact that, although the configuration generated by RMC offers an excellent match with neutron diffraction data, the configuration generated by RMC is unstable and, hence, fairly unrealistic. This arises from the fact that, as mentioned above, many configurations (stable or not) can exhibit the same pair distribution function, highlighting the intrinsic limitations of RMC refinement.

In contrast, we observe that FEAR yields a potential energy that is significantly lower than that offered by MQ-MD (see **Fig. 5a**), in agreement with previous findings [26]. This is significant since it signals that, although FEAR and MQ-MD relies on the same interatomic forcefield, imposing a structural constraint (i.e., the neutron PDF) allows the structure to reach lower energy states. This suggests that the RMC refinement steps performed in between each energy minimization allow the system to jump over some large energy barriers and, thereby, reach some energy basins that would be inaccessible during the limited timescales of MD simulations. Overall, these results demonstrate that FEAR refinement can generate some glass structures that simultaneously exhibit unprecedented level of agreement with experimental data and energetic stability.

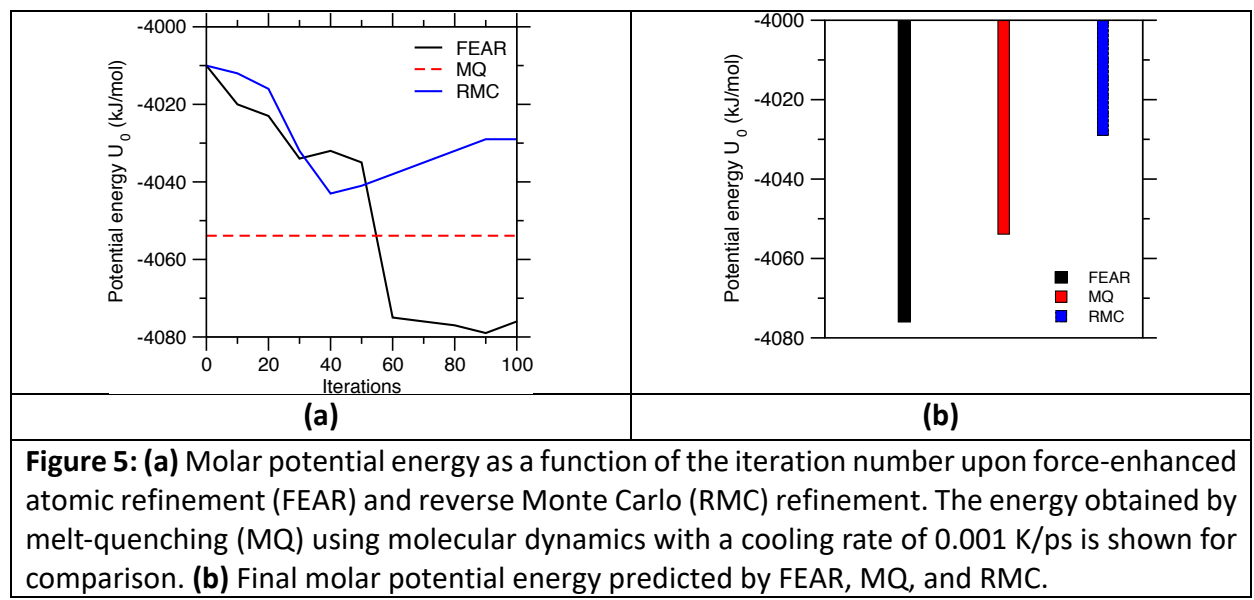

**Figure 5: (a)** Molar potential energy as a function of the iteration number upon force-enhanced atomic refinement (FEAR) and reverse Monte Carlo (RMC) refinement. The energy obtained by melt-quenching (MQ) using molecular dynamics with a cooling rate of 0.001 K/ps is shown for comparison. **(b)** Final molar potential energy predicted by FEAR, MQ, and RMC.

To further appreciate the magnitude of the variations in $R_\chi$ and potential energy yielded by FEAR and RMC, we compare them to the variations resulting from the use of different cooling rates in MQ-MD simulations. **Figure 6a** first shows the evolution of $R_\chi$ as a function of a function of the cooling rate (as computed from the MQ-MD simulations). As expected, we find that $R_\chi$ decreases upon decreasing cooling rate [9], which indicates that the PDF of the glass structures generated by MQ-MD gradually converges toward the experimental neutron diffraction data as the cooling rate decreases. Nevertheless, independently of the cooling rate, the $R_\chi$ coefficient of the glass structures simulated by MQ-MD systematically remains significantly higher than those obtained by RMC and FEAR. Based on cooling-rate dependence predicted by mode-coupling theory [48], the $R_\chi$ are fitted by a power law and extrapolated to lower cooling rates. Notably, this rough extrapolation of the MD-derived $R_\chi$ coefficients as a function of the cooling rate (assuming a power law dependence, see **Fig. 6a** [9,48]) suggests that a cooling rate of about 100 K/s (i.e., $10^{-10}$ K/ps, which is fairly comparable to experimental cooling rates) would be needed for a melt-quench MD simulation to yield a $R_\chi$ coefficient that is comparable to that offered by FEAR. Such a simulation would roughly require a CPU time of 100 million years to complete. This highlights the clear advantage of FEAR refinement over conventional melt-quenching MD simulations.

**Figure 6b** shows the evolution of the molar potential energy of the MD-based simulated glasses as a function of the cooling rate. As expected, the potential energy decreases upon decreasing cooling rate [9], which indicates that the system becomes more stable and reaches deeper states within the energy landscape (or achieves lower fictive temperatures) [49,50]. We note that the potential energy offered by RMC is significantly higher than that offered by MQ-MD, even for a very high cooling rate of 100 K/ps. In turn, we find that the magnitude of the decrease in potential energy provided by FEAR as compared to a 0.001 K/ps MQ-MD simulation (i.e., about 20 kJ/mol) is comparable to that observed upon a variation of 5 orders of magnitude of the cooling rate (i.e., from 100 to 0.001 K/ps, see **Fig. 6b**).

Finally, we focus on the evolution of glass density. As shown in **Fig. 6c**, the glass density predicted by MQ-MD simulations expectedly tends to increase upon decreasing cooling rate. This echoes the fact that glasses usually become more compact upon slower cooling rate (with the notable exception of pure silica) [14,50]. However, as the cooling rate decreases, we note that MQ-MD simulations tend to overestimate the density of sodium silicate observed experimentally, i.e., 2.45 g/cm$^3$ [44,51]. The origin of this discrepancy is unclear and might be a spurious effect of the melt-quenching procedure (for instance, the empirical forcefield used herein may not be accurate at high-temperature, i.e., when the atoms are far from equilibrium). In contrast, we find that FEAR yields a glass density that shows an excellent match with experimental data (see **Fig. 6d**). This suggests that the fact that the MQ-MD simulation overestimates the glass density is not simply limitation of the empirical forcefield (at room temperature). Note that, in the pure RMC simulations, the density is fixed to be equal to the experimental value—so that RMC simulations cannot be used to predict the glass density. Again, these results suggest that the combination of energy minimization and experimental constraints can overcome the limitations of MQ-MD and RMC simulations.

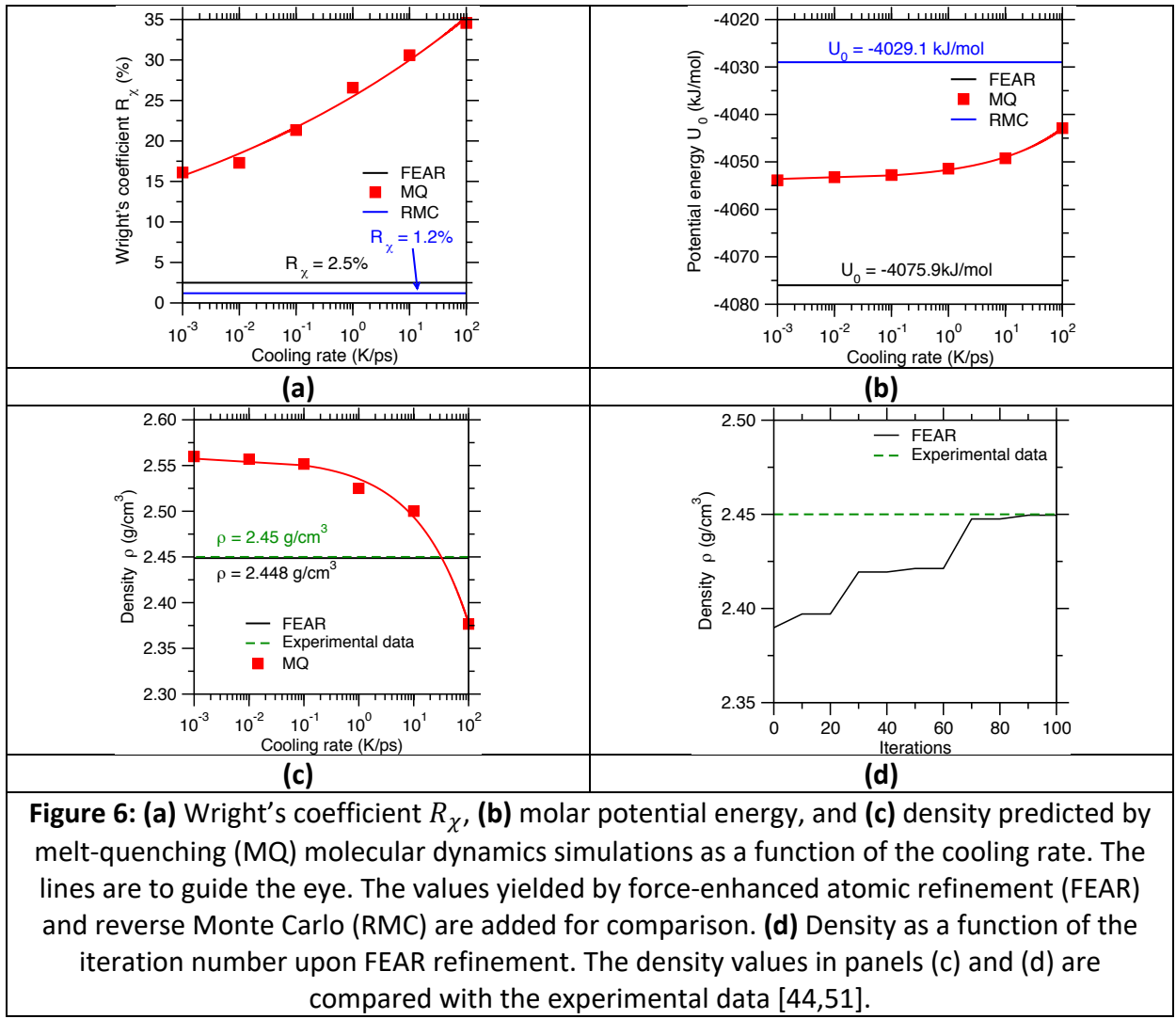

**Figure 6: (a)** Wright's coefficient $R_\chi$, **(b)** molar potential energy, and **(c)** density predicted by melt-quenching (MQ) molecular dynamics simulations as a function of the cooling rate. The lines are to guide the eye. The values yielded by force-enhanced atomic refinement (FEAR) and reverse Monte Carlo (RMC) are added for comparison. **(d)** Density as a function of the iteration number upon FEAR refinement. The density values in panels (c) and (d) are compared with the experimental data [44,51].

### 3.3 *Short-range order signatures of the force-enhanced refinement*

We now explore how the level of improvement offered by FEAR refinement (in terms of $R_\chi$ and energetic stability) manifests itself in the short-range order (<3 Å) structure of the simulated sodium silicate glass. **Figure 7** shows all the partial PDFs obtained by MQ-MD, RMC, and FEAR. We first note that the positions of the Si–O and O–O peaks around 1.6 and 2.6 Å, respectively, remain fairly unaffected, although FEAR and RMC appear to predict a sharper peak, that is, a more ordered local environment for Si atoms (see **Figs. 7a** and **7b**). In contrast, we note the Si–Si peak around 3.1 Å predicted by FEAR and RMC shifts toward higher distances compared to MQ-MD (see **Fig. 7c**). This behavior echoes the peak shift observed upon decreasing cooling rate in MQ-MD simulations [9]. The fact that the average Si–Si distance increases upon FEAR refinement while the Si–O bond distance remains constant suggests an opening of the Si–O–Si angle (see below). We then note that that the Na–O PDF remains largely unaffected (see **Fig. 7d**), which suggests that what the local environment of the Na atoms only weakly depends on the simulation technique that is used. More significant variations are observed in the Na–Na PDF (see **Fig. 7e**). Specifically, we find that the peak intensity predicted by RMC is lower than those offered by MQ-

Page 11

MD and FEAR, which suggests that RMC underestimates the degree of modifier clustering [9]. Finally, we note that both FEAR and RMC simulations predict a Si–Na PDF that is more asymmetric than that predicted by MQ-MD (see **Fig. 7f**).

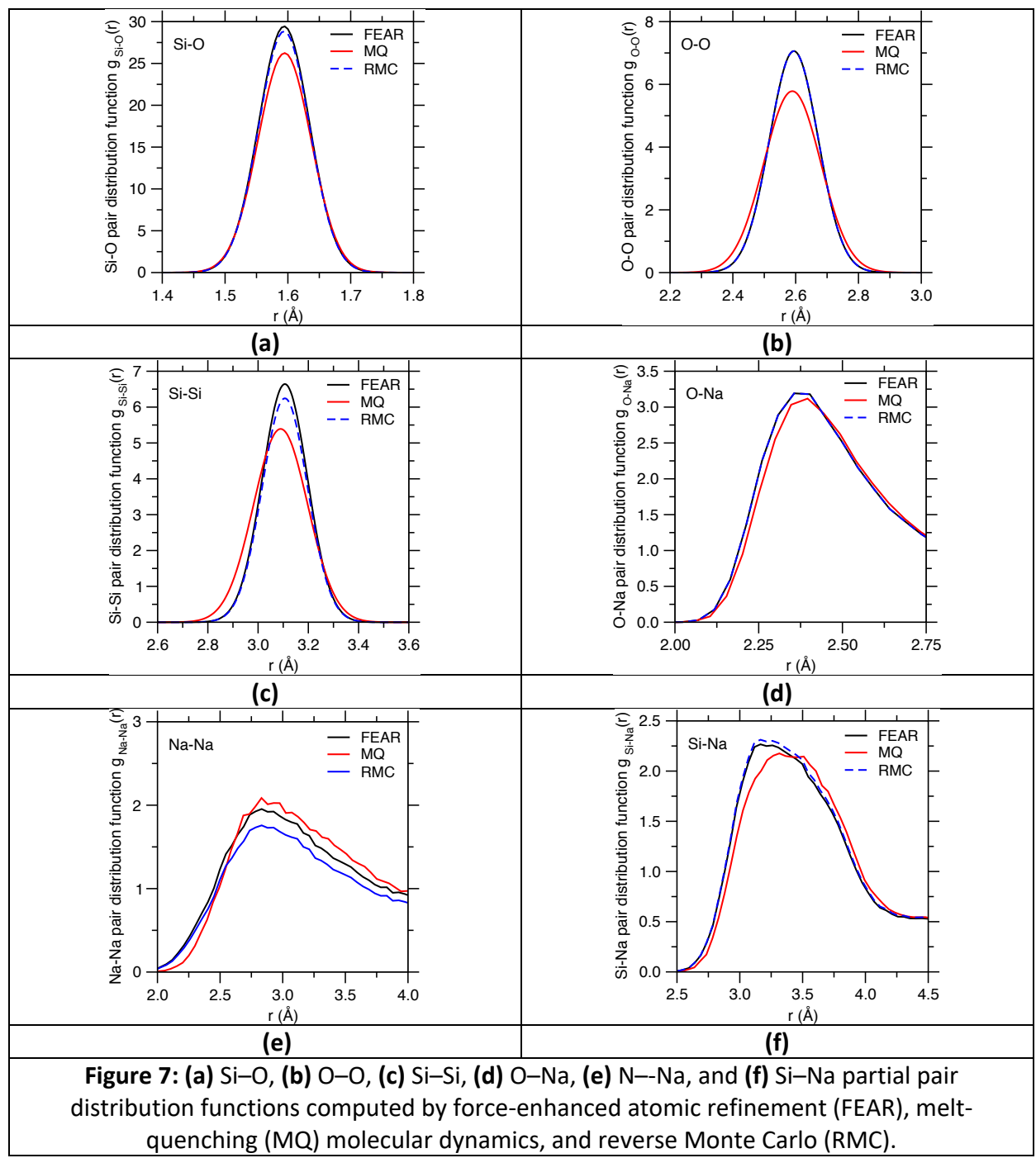

**Figure 7: (a)** Si–O, **(b)** O–O, **(c)** Si–Si, **(d)** O–Na, **(e)** N–-Na, and **(f)** Si–Na partial pair distribution functions computed by force-enhanced atomic refinement (FEAR), melt-quenching (MQ) molecular dynamics, and reverse Monte Carlo (RMC).

Then we direct our attention to the angular environment of each element. **Figure 8** shows the O–Si–O (intratetrahedral) and Si–O–Si (intertetrahedral) partial bond angle distributions (PBADs) offered by FEAR, MD, and RMC. First, we note that the average O–Si–O angle remains unaffected, with a PBAD centered around 109°, which is in agreement with the tetrahedral environment of the Si atoms. However, we observe that the O–Si–O PBAD becomes sharper upon FEAR refinement, which denotes the existence of a more ordered angular environment for the Si atoms. This echoes the fact that the O–Si–O PBAD computed by MD simulations also becomes sharper upon decreasing cooling rate [9]. Second, we observe that that the Si–O–Si PBAD becomes sharper and shifts toward higher angle values upon FEAR refinement. The behavior is supported by RMC simulations and denotes an opening of the intertetrahedral angle. This also echoes the fact that the average Si–O–Si angle computed by MD simulations increases upon decreasing cooling rate [9]. Overall, this suggests that more relaxed glassy silicate structures are associated with more open intertetrahedral angles [9].

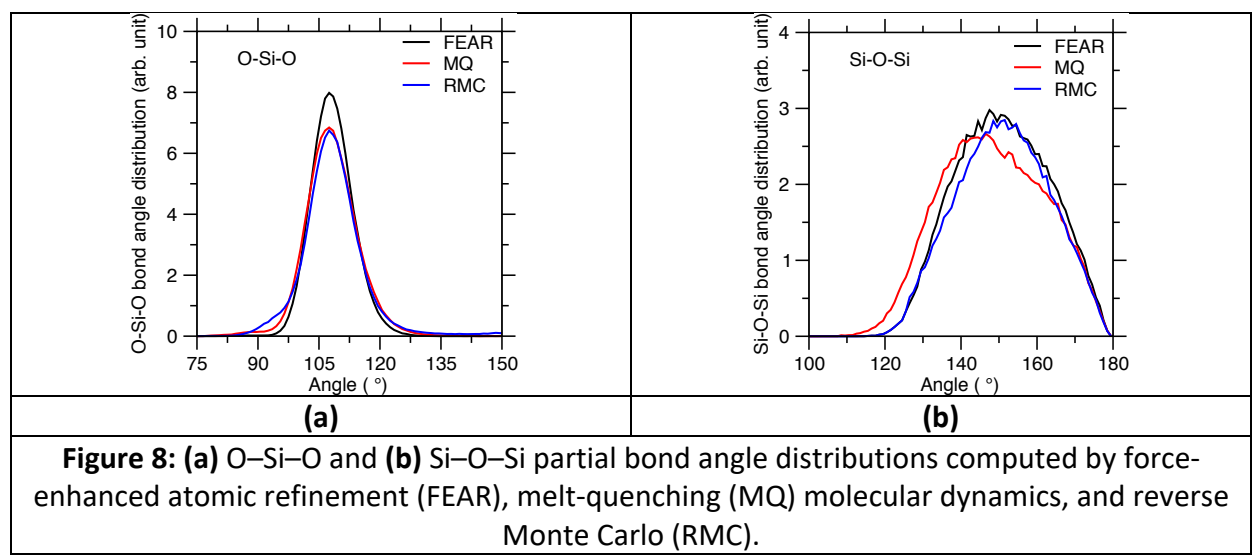

**Figure 8: (a)** O–Si–O and **(b)** Si–O–Si partial bond angle distributions computed by force-enhanced atomic refinement (FEAR), melt-quenching (MQ) molecular dynamics, and reverse Monte Carlo (RMC).

### 3.4 *Topological signatures of the force-enhanced refinement*

We now investigate the variation in network connectivity upon refinement. Although FEAR, MD, and RMC yield similar coordination numbers (e.g., 4 for Si) and fraction of non-bridging oxygen (NBO, i.e., each Na creates 1 NBO), we observe some variations in the interconnectivity of the SiO$_4$ tetrahedral polytopes. Such connectivity is captured by the $Q_n$ distribution, wherein a $Q_n$ species denotes an SiO$_4$ tetrahedral unit that is connected to $n$ other Si atoms and, hence, that comprises $4 - n$ NBOs.

Due to the high cooling rates they rely on, melt-quenching MD simulations typically yield glass structures that are more disordered than those observed experimentally [13]. Consequently, MD simulations tend to predict glass structures wherein the NBO are more randomly attached to the Si atoms. As such, MD simulations typically yield $Q_n$ distributions that agree well with the predictions from a random model [21,28,52], but, in turn, do not match with experimental data [53]. Although lower cooling rates tend to improve the agreement between MD simulations and experimental data [9], we observe that, even with a slow cooling rate of 0.001 K/ps (as used herein), MD simulations yield a fairly random $Q_n$ distribution. Such randomness manifests itself by the fact that MD (i) underestimates the fraction of $Q_3$ units and (ii) overestimates the fraction of $Q_0$ defects (see Ref. [9]).

As shown in **Fig. 9**, FEAR refinement partially addresses these two points. First, both FEAR and RMC yield a significantly lower fraction of $Q_0$ defects—and such units virtually disappear upon FEAR refinement (see **Fig. 9b**), in agreement with experimental data [9]. Second, we find that, although RMC predicts a fraction of $Q_3$ units that is comparable to MD, FEAR yields an increased fraction of $Q_3$ units. The preferential formation of $Q_3$ units is also observed in MD simulations upon decreasing cooling rate [9] and has been explained by the fact that such units are isostatic and exhibit minimum internal stress [54]. Overall, this shows that FEAR yields a glass topology that is more ordered (i.e., less random) than those predicted by MD and RMC.

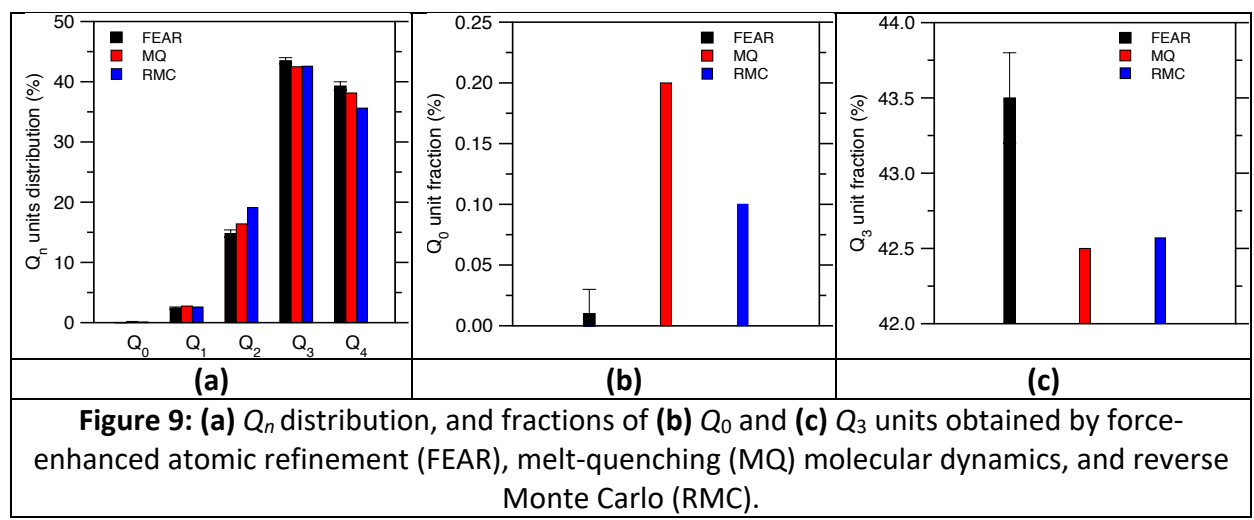

**Figure 9:** **(a)** $Q_n$ distribution, and fractions of **(b)** $Q_0$ and **(c)** $Q_3$ units obtained by force-enhanced atomic refinement (FEAR), melt-quenching (MQ) molecular dynamics, and reverse Monte Carlo (RMC).

### 3.5 *Medium-range order signatures of the force-enhanced refinement*

Finally, we investigate how FEAR refinement manifests itself in the medium-range order structure of the simulated sodium silicate glass. In disordered silicate systems, the medium-range order is primarily captured by the ring size distribution—wherein silicate rings are defined as the shortest closed paths made of Si–O bonds within the atomic network [55]. Here, the ring size distribution of each configuration is computed by using the RINGS package [56].

**Figure 10** shows that ring size distribution computed by FEAR, MD, and RMC. We first note that the ring size distribution offered by MD is in agreement with previous works, being centered around 5-to-6-membered rings [10,40]—wherein the ring size is expressed in terms of the number of Si atoms it comprises [57]. Interestingly, we then observe that both FEAR and RMC yield a sharper ring size distribution than that predicted by MD (see **Fig. 10a**). Specifically, both FEAR and RMC predict a lower fraction of small rings (i.e., 4-membered and smaller, see **Fig. 10b**). In particular, 2-membered (i.e., edge-sharing units) and 3-membered rings are virtually absent from the glass structure after FEAR refinement. Similarly, both FEAR and RMC predict a lower fraction of larger rings (i.e., 7-membered and larger, see **Fig. 10c**). In turn, both FEAR and RMC predict a higher fraction of 6-membered rings than MD. These results echo the fact that the fractions of both small and large rings predicted by MD tend to decrease upon decreasing cooling rate [40]. This is also in agreement with the fact that small rings are over-constrained and, hence, experience some internal stress [40,55]. Overall, these results suggest that, when properly relaxed, sodium silicate glasses exhibit a more ordered medium-range order structure (i.e., sharper ring size distribution) than suggested by MD simulations.

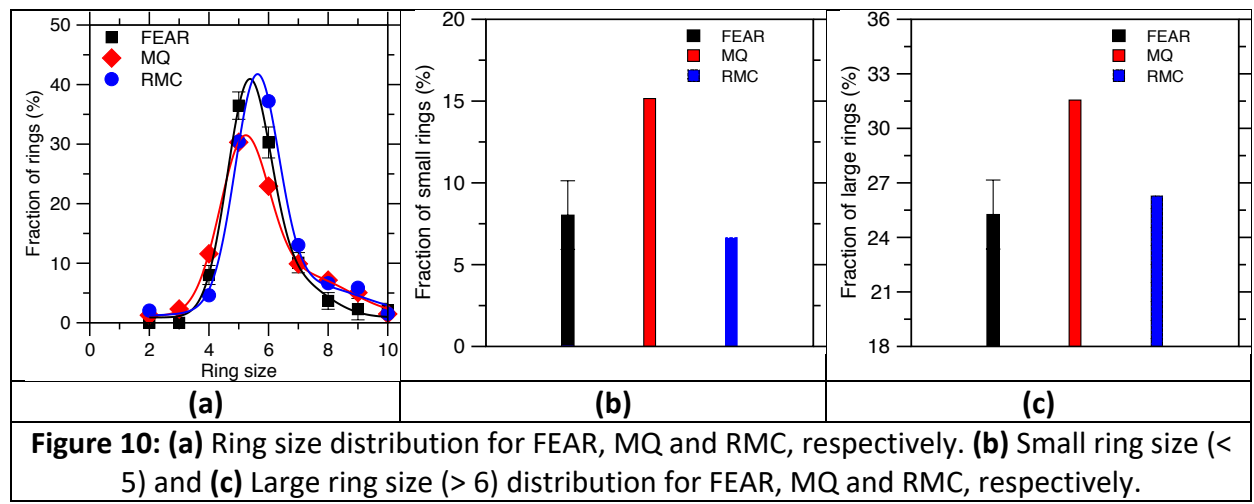

**Figure 10: (a)** Ring size distribution for FEAR, MQ and RMC, respectively. **(b)** Small ring size (< 5) and **(c)** Large ring size (> 6) distribution for FEAR, MQ and RMC, respectively.

## 6. Conclusions

Overall, FEAR refinement yields a sodium silicate glass structure that simultaneously exhibits enhanced agreement with diffraction data and higher energetic stability as compared to configurations generated by MD or RMC. As such, FEAR offers an unprecedented description of the atomic structure of sodium silicate glasses. We find that the enhanced thermodynamic stability offered by FEAR primarily finds its roots in the $Q_n$ and ring size distributions of the network. Our findings thus suggest that the structure of silicate glasses is more ordered, less random than previously suggested by MD simulations. Overall, this study establishes FEAR as a promising tool to explore the complex, largely hidden atomic structure of disordered solids.

**Acknowledgements**

The authors acknowledge financial support for this research provided by the National Science Foundation under Grant No. 1928538 and the International Cooperation on Scientific and Technological Innovation Programs of BGRIMM under Grant No. 2017YFE0107000. M.M.S. acknowledges funding from the Independent Research Fund Denmark (Grant No. 7017-00019).